\documentclass[11pt]{article}

\usepackage{amsmath}
\usepackage{amssymb}
\usepackage{graphicx}

\begin{document}

\begin{center}
\bf\Large RadioAstron gravitational redshift experiment: \\status update
\end{center}

\begin{center}
D.~A.~Litvinov$^{1,2}$,
U.~Bach$^{3}$,
N.~Bartel$^4$,
K.~G.~Belousov$^2$,
M.~Bietenholz$^{4,5}$,
A.~V.~Biriukov$^2$,
G.~Cim\'o$^{6,7}$,
D.~A.~Duev$^{1,8}$,
L.~I.~Gurvits$^{6,9}$,
A.~V.~Gusev$^1$,
R.~Haas$^{10}$,
V.~L.~Kauts$^{2,11}$,
B.~Z.~Kanevsky$^{2}$,
A.~V.~Kovalenko$^{12}$,
G.~Kronschnabl$^{13}$,
V.~V.~Kulagin$^1$,
M.~Lindqvist$^{10}$,
G.~Molera~Calv\'es$^{6,14}$,
A.~Neidhardt$^{15}$,
C.~Pl\"otz$^{13}$,
S.~V.~Pogrebenko$^6$,
N.~K.~Porayko$^{1,3}$,
V.~N.~Rudenko$^1$,
A.~I.~Smirnov$^{2}$,
K.~V.~Sokolovsky$^{1,2,16}$,
V.~A.~Stepanyants$^{17}$,
J.~Yang$^{10}$,
M.~V.~Zakhvatkin$^{17}$
\end{center}

{\small

\noindent$^1$~Sternberg Astronomical Institute, Lomonosov Moscow State University,
Universitetsky pr.~13, 119991 Moscow, Russia, e-mail: litvirq@yandex.ru

\noindent$^2$ Astro Space Center, Lebedev Physical Institute, Profsoyuznaya 84/32, 117997 Moscow, Russia

\noindent$^{3}$ Max-Planck-Institut f\"ur Radioastronomie, Auf dem H\"ugel 69, 53121 Bonn, Germany

\noindent$^4$ York University, Toronto, Ontario M3J 1P3, Canada

\noindent$^5$ Hartebeesthoek Radio Observatory, P.O. Box 443, Krugersdorp 1740, South Africa

\noindent$^6$ Joint Institute for VLBI ERIC, PO Box 2, 7990 AA Dwingeloo, The Netherlands

\noindent$^{7}$ ASTRON, the Netherlands Institute for Radio Astronomy, PO 2, 7990 AA Dwingeloo, The~Netherlands

\noindent$^8$ California Institute of Technology, Pasadena, CA 91125, USA

\noindent$^{9}$ Department of Astrodynamics and Space Missions, Delft University of Technology, 2629~HS~Delft, The Netherlands

\noindent$^{10}$ Department of Earth and Space Sciences, Chalmers University of Technology, Onsala Space Observatory, 439 92 Onsala, Sweden

\noindent$^{11}$ Bauman Moscow State Technical University, 2-ya Baumanskaya 5, 105005 Moscow, Russia

\noindent$^{12}$ Pushchino Radio Astronomy Observatory, 142290 Pushchino, Russia

\noindent$^{13}$ Federal Agency for Cartography and Geodesy, Sackenrieder Str.~25, D-93444 Bad K\"otzting, Germany

\noindent$^{14}$ Aalto University, School of Electrical Engineering, Department of Radio Science and Engineering, 02120 Espoo, Finland

\noindent$^{15}$ Technical University of Munich, Geodetic Observatory Wettzell, Sackenrieder Str.~25, D-93444 Bad K\"otzting, Germany

\noindent$^{16}$ Institute   of   Astronomy,    Astrophysics,    Space   Applications  and  Remote  Sensing,  National  Observatory  of  Athens, Vas. Pavlou \& I. Metaxa, GR-15 236 Penteli, Greece

\noindent$^{17}$ Keldysh Institute for Applied Mathematics, Russian Academy of Sciences, Miusskaya sq. 4, 125047 Moscow, Russia

} %\small

\begin{abstract}
A test of a cornerstone of general relativity, the gravitational redshift effect, is currently being conducted with the RadioAstron spacecraft, which is on a highly eccentric orbit around Earth. Using ground radio telescopes to record the spacecraft signal, synchronized to its ultra-stable on-board H-maser, we can probe the varying flow of time on board with unprecedented accuracy. The observations performed so far, currently being analyzed, have
already allowed us to measure the effect with a relative accuracy of $4\times10^{-4}$. We expect to reach $2.5\times10^{-5}$ with additional observations in 2016, an improvement of almost a magnitude over the 40-year old result of the GP-A mission.
\vspace{.3cm}

\end{abstract}

\section{Introduction}\label{aba:sec1}
Quantum theory (QT) and general relativity (GR) are the two pillars of modern physics. However, they are incompatible.
Theoretical attempts to unify QT and GR lead to violations of GR and, in
particular, the
equivalence principle (EP). It is hard to estimate the level at which EP may be violated. Therefore
equivalence principle tests are considered to ``stand out as our deepest possible probe of new physics'' \cite{damour-2012}. We intend to test the EP with RadioAstron.

According to the Einstein  equivalence principle, an electromagnetic wave of frequency $f$, propagating in a region of space where the gravitational potential is not
constant, experiences a gravitational frequency shift:
\begin{equation}
   {\Delta f_\mathrm{grav} \over f} = { \Delta U \over c^{2} },
\label{eq:main}  
\end{equation}
where $\Delta U$ is the
gravitational potential difference between the measurement points and $c$ is the speed of light \cite{misner-thorne-wheeler}. Any violation of Eq. (\ref{eq:main}) in an experiment with
two identical atomic frequency standards may be parameterized
as
\begin{equation}
{\Delta f_\mathrm{grav} \over f} = {\Delta U \over
c^2} (1+\varepsilon),
\label{eq:main-violated}
\end{equation}where the violation parameter, $\varepsilon$, may depend on element  composition of the gravitational field sources and on the specific
kind of quantum transition exploited by
the frequency standards.  It is generally agreed that the best test of Eq. (\ref{eq:main})
to date was performed with the NASA-SAO Gravity Probe A
(GP-A) \cite{vessot-1980-prl} mission 40 years ago which  measured $\varepsilon=(0.05\pm1.4)\times10^{-4}$,
giving the accuracy $\delta\varepsilon=1.4\times10^{-4}$.
The gravitational potential modulation experienced by RadioAstron is comparable
to that of GP-A: $\Delta U/c^2 \sim 3\times10^{-10}$. The better stability of the RadioAstron on-board H-maser and the possibility of repeating observations
promise a factor of $\sim$~10 improvement on the GP-A result.

Testing the gravitational redshift effect has recently become
an active field of experimental research. The experiment with Galileo~5~\&~6 navigational satellites is expected to probe the effect with (3--4)$\times10^{-5}$ accuracy \cite{delva-2015}. The specialized ACES mission \cite{aces}, to be launched in 2017, is expected
to achieve $2\times10^{-6}$.

\section{Outline of the Experiment}
In the gravitational redshift experiment with RadioAstron we detect the frequency change of the RadioAstron's on-board H-maser due to gravitation by comparing it, by means of radio links,  with an H-maser at a
ground station. Either one 
of the mission's dedicated tracking stations (TS), Pushchino or Green Bank, or a ground
 radio telescope (GRT) equipped with a
8.4 or 15 GHz receiver
 and appropriate data acquisition instrument may be used for receiving the spacecraft signal. The frequency variation due to the small gravitational frequency shift ($\Delta f/f\sim10^{-10}$) needs to be separated from a number of other effects influencing the signal sent from the spacecraft to the ground station \cite{vessot-1980-prl}:

\begin{align}
{\Delta f_{1\mathrm{w}}} &= 
f \left(
 - {\dot D \over c} 
- {v_\mathrm{s}^2 - v_\mathrm{e}^2 \over 2 c^2}
+ { (\mathbf v_\mathrm{s} \cdot \mathbf n)^2
- (\mathbf v_\mathrm{e} \cdot \mathbf n) \cdot (\mathbf v_\mathrm{s} \cdot \mathbf n)
  \over c^2 }
\right)  \nonumber \\
&+ {\Delta f_\mathrm{grav} }%\over f}
+ {\Delta f_\mathrm{ion} }%\over f} 
+ {\Delta f_\mathrm{trop} } 
+ {\Delta f_0 }%\over f}
+ O\left({v\over c}\right)^3,
\label{eq:one-way}
\end{align}

\noindent
where ``$\mathrm{1w}$'' stands for ``1-way'' (space
to ground  link), $\mathbf v_\mathrm{s}$ and $\mathbf v_\mathrm{e}$ are the velocities of the spacecraft and
the ground station (in an Earth-centered inertial reference
frame), $\dot D$ is the radial velocity of the
spacecraft relative to the ground station, $\Delta f_\mathrm{grav}$ is the
gravitational redshift, $\mathbf n$ is a unit vector in the direction opposite to that of signal propagation, $\Delta f_\mathrm{ion}$  and $\Delta f_\mathrm{trop}$ are the ionospheric and tropospheric shifts, and $\Delta f_0$ is the frequency bias
between the ground and space H-masers.

There are two major problems in using Eq.~(\ref{eq:one-way}) to determine $\Delta f_\mathrm{grav}$ directly, at
least for RadioAstron. The first is caused by the unknown frequency bias, $\Delta f_0$, which cannot be determined after launch without making
use of Eq.~(\ref{eq:main}). We solve
this problem
by measuring only the variation of the gravitational effect and
taking into account the bias drift.
The second problem  is  that the  nonrelativistic Doppler shift, $-\dot D/c$, cannot be calculated accurately enough from the
available spacecraft state vector data. We solve this problem with the help of the frequency measurements
of the 2-way link, which let us cancel the
nonrelativistic Doppler term:
\begin{align}
\Delta f_{1\mathrm{w}} - {1\over 2}\Delta f_{2\mathrm{w}} 
&=
\Delta f_\mathrm{grav} + f \left(
- {|\mathbf v_\mathrm{s}^2 - \mathbf v_\mathrm{e}^2| \over 2 c^2}
+ {\mathbf a_\mathrm{e} \cdot \mathbf n_{} \over c} \Delta t 
 \right) + O(v/c)^3,
\label{eq:gpa-compensation-scheme}
\end{align}
\noindent
where $\mathbf a_e$ is the ground station acceleration and $\Delta t$ is the
signal light time. (Eq.~\eqref{eq:gpa-compensation-scheme} is relevant for a TS, similar but more complex equation
holds for the case of the 2-way link signal received by a nearby GRT.)  The idea of the compensation scheme based on Eq.~(\ref{eq:gpa-compensation-scheme}) was first implemented in the GP-A experiment.
For RadioAstron, however, 
%\noindent
it is not directly applicable due to impossibility of the 1- and 2-way links to be using different reference signals  simultaneously (Fig.~1).
Nevertheless, two options for realizing the compensation scheme (\ref{eq:gpa-compensation-scheme}) with RadioAstron are available.

The first option requires interleaving the \mbox{1-way}   (Fig.~1a) and 2-way   (Fig.~1b)  modes \cite{litvinov-2015-journees}. The data
recorded by GRTs (and the TS) contain only one kind of signal at any given time.  
However, if  the switching
cycle is short enough ($\sim$ 4~min at 8.4~GHz) we are able to interpolate into the gaps with an error of $\Delta f/f\sim5\times10^{-15}$. Thus we obtain quasi-simultaneous
frequency measurements of both kinds and can apply the compensation
scheme (\ref{eq:gpa-compensation-scheme}) to them directly.

\def\figsubcap#1{\par\noindent\centering\footnotesize(#1)}
\begin{figure}[h]%
\begin{center}
  \parbox{1.4in}{\includegraphics[width=1.4in]{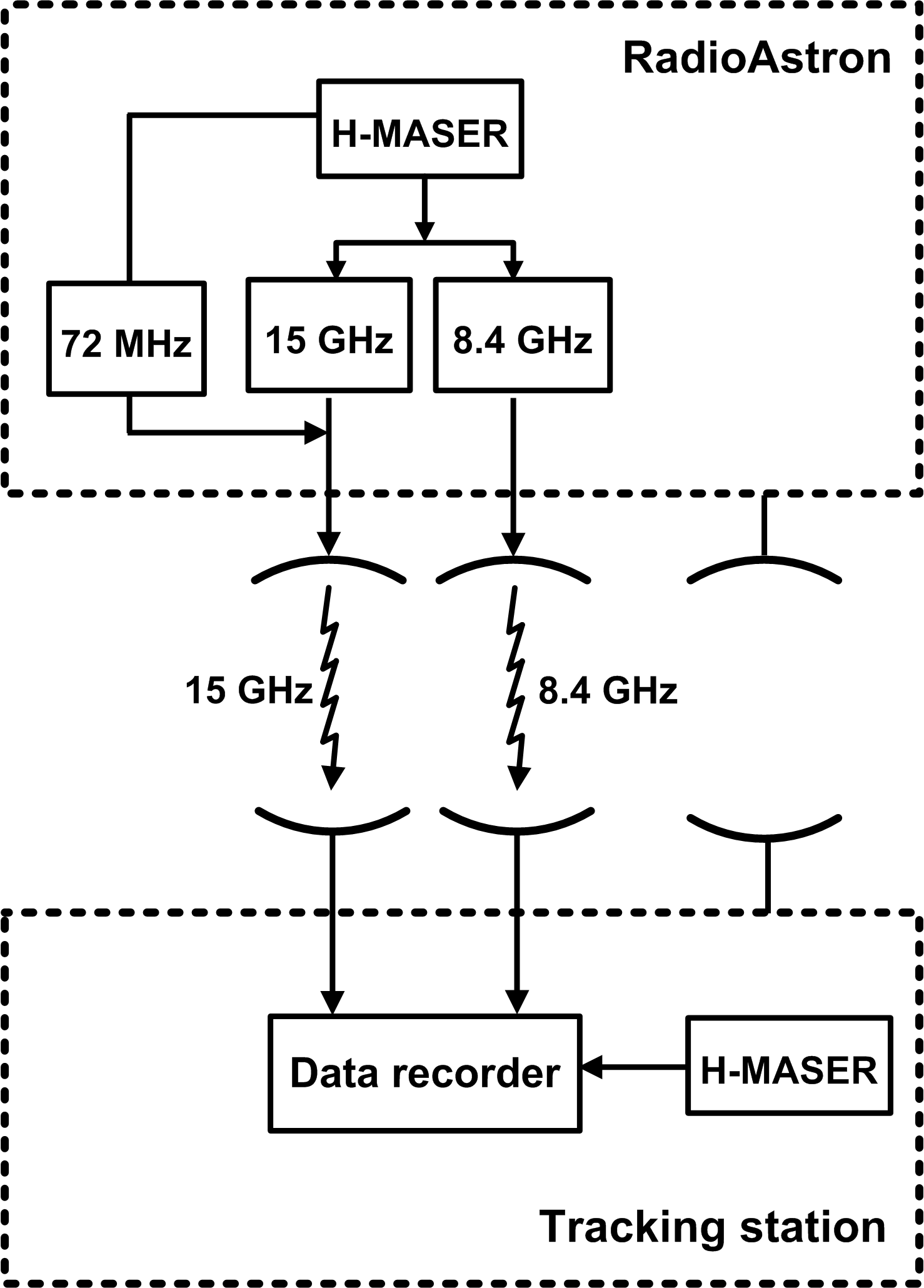}\figsubcap{a}}
  \hspace*{10pt}
  \parbox{1.4in}{\includegraphics[width=1.4in]{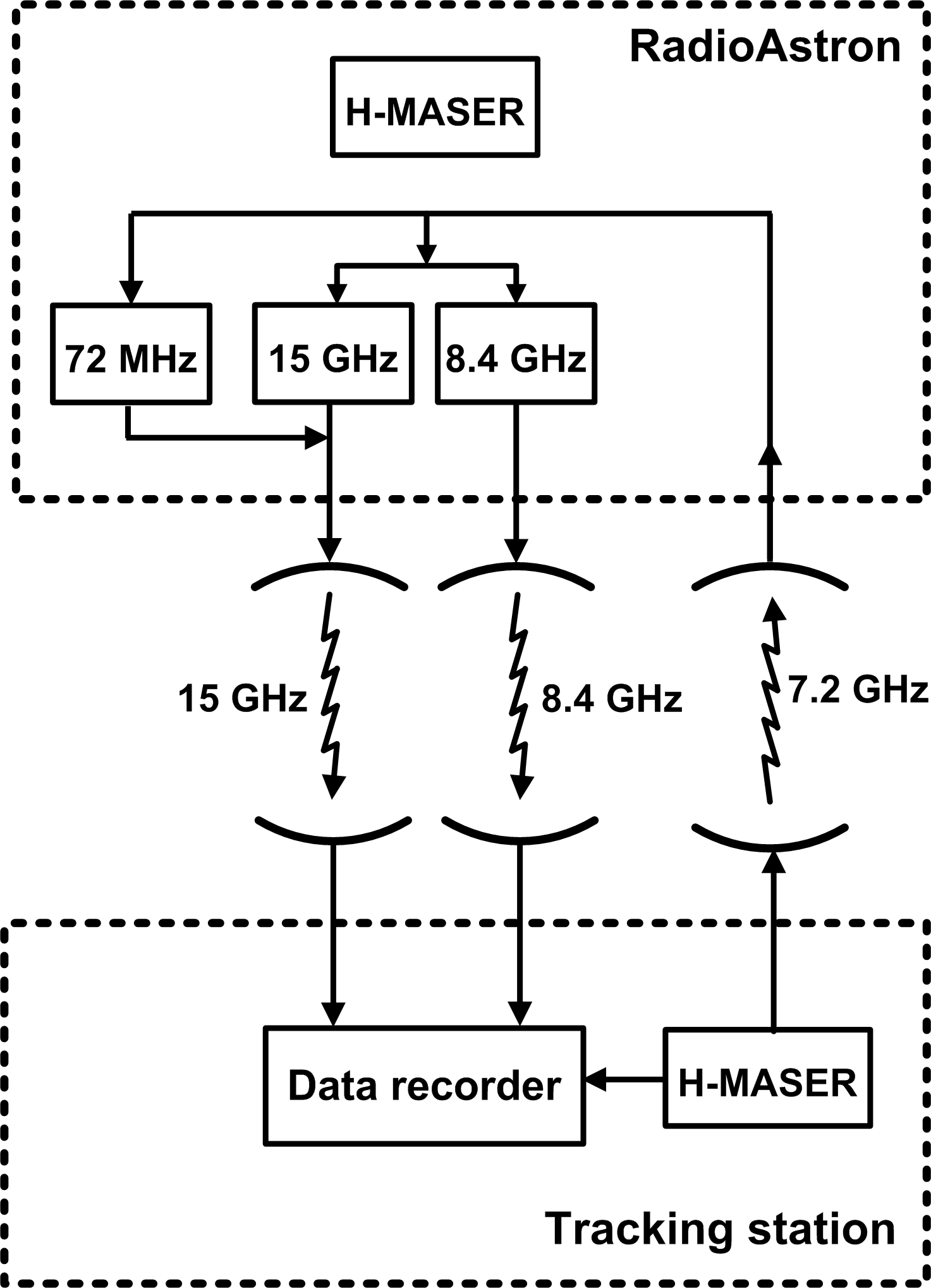}\figsubcap{b}}
  \hspace*{10pt}
  \parbox{1.4in}{\includegraphics[width=1.4in]{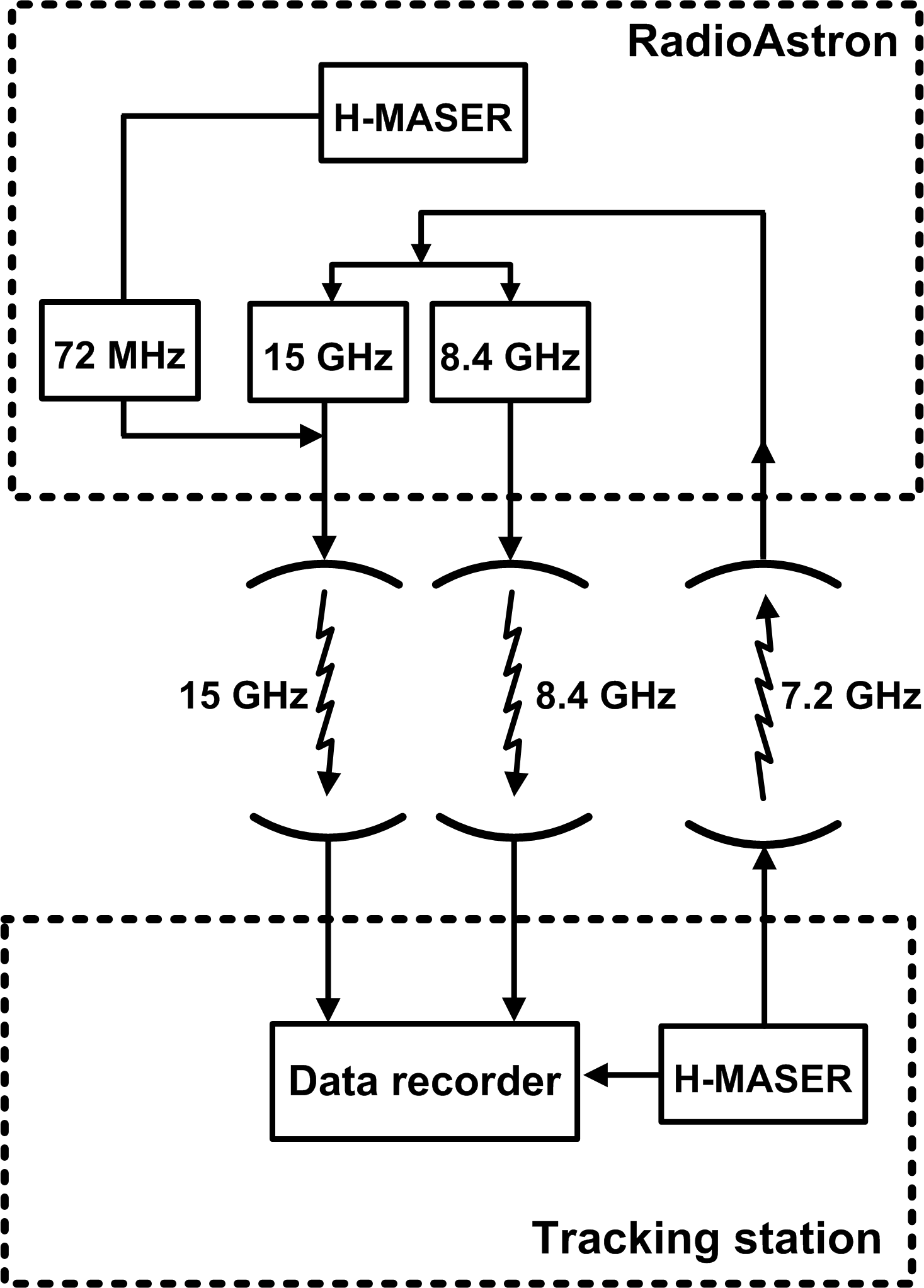}\figsubcap{c}}
  \caption{On-board hardware  synchronization modes. (a) ``H-Maser''. (b) ``Coherent''. (c) ``Semi-Coherent''. Note that, due to a fixed architecture of the RadioAstron on-board radio system, the 8.4 GHz tone and the carrier of the 15 GHz data link must use one and the same reference signal at any
given time:
either from the on-board H-maser or from the tracking station uplink.}%
  \label{fig1.2}
\end{center}
\end{figure}

\vspace{-.2cm}

The second option involves
recording the 15~GHz  data link signal in the  ``Semi-Coherent'' mode of the on-board scientific and radio equipment (Fig.~1c). In this mode the  7.2~GHz
uplink tone, the 8.4 GHz downlink tone and the 15 GHz data downlink
carrier are phase-locked to the ground H-maser signal, while the modulation frequency
  of the data downlink is phase-locked to the on-board H-maser \cite{biriukov-2014-ar}.
We do not give a detailed account of this approach here because it will
not likely be used in the observations due to technical reasons. The basic data processing algorithms in each approach are  those developed originally for PRIDE (Planetary Radio Interferometry and Doppler Experiment) \cite{duev-2012-aa}.

{
\renewcommand{\arraystretch}{1.02}

\footnotesize

\vspace{.5cm}
%\noindent
\hspace{.7cm}
Table 1. Error budget of the RadioAstron gravitational redshift experiment
\begin{center}
\vspace{-.2cm}

\centering
\nobreak
\begin{tabular}{lr}
\hline
Random errors in ${\Delta f \over f}$:  \\
\quad Frequency instability $\sigma_y(\tau$ = 1000 s):\\
\quad\quad 1-way signal
&
$2\cdot10^{-14}$
\\
\quad\quad 2-way signal
&
$3\cdot10^{-14}$
\\
\quad Interpolation error
&
$5\cdot10^{-15}$
\\
\quad Uncancelled tropospheric noise
&
$2\cdot10^{-15}$
\\
\quad Uncancelled ionospheric noise
&
$1\cdot10^{-15}$
\\
\quad Net random error  $\delta_r{\Delta f \over f}$:\\
\quad\quad single experiment
&
$4\cdot10^{-14}$
\\
\quad\quad 30 experiments 
&
$8\cdot10^{-15}$
\\
\hline
Systematic errors in ${\Delta f \over f}$:\\
\quad Residual space and ground clock drift over single experiment 
&
$1\cdot10^{-15}$
\\
\quad Redshift and 2nd-order Doppler prediction errors due to   orbit inaccuracy     
&
$1\cdot10^{-15}$
\\
\quad Tropospheric and ionospheric bias
&
$2\cdot10^{-15}$
\\
\quad Net systematic error $\delta_s{\Delta f \over f}$
&
$2\cdot10^{-15}$
\\
\hline
Total error (random + systematic) $\delta{\Delta f \over f}$:
\\
\quad single experiment
&
$4\cdot10^{-14}$
\\
\quad 30 experiments
&
$8\cdot10^{-15}$
\\
\hline
Average variation of $\Delta U/c^2$ 
&
$3.0\cdot10^{-10}$
\\
Predicted experiment accuracy  $\delta(\Delta f / f) \over \Delta U/c^2$:
\\
\quad single experiment &
$1.0\cdot10^{-4}$
\\
\quad 30 experiments &
$2.5\cdot10^{-5}$
\\
\hline
%}
\end{tabular}
\end{center}
}

\vspace{.4cm}

Based on the  error budget of the experiment using the interleaved measurements  approach (Table 1), we expect
the accuracy of the redshift test to reach 
\begin{equation}
\delta\varepsilon\sim 2.5\times10^{-5},
\end{equation}   
which is almost an order of magnitude better than the result of the GP-A mission. \,This value \;\,takes \;\,into \;\,account \;\,a \;number \;of \;factors\; not covered in the
discussion above, such as ionospheric and tropospheric correction of GRT
data using on-site GPS measurements, uncertainty of
orbit reconstruction, etc.

\section{Current Status}
\setlength{\parfillskip}{0pt}
A total of 6~experiments have been performed in the period from October to December
2015, all
using the interleaved measurements approach.
Half of them were performed with the National Radio Astronomy Observatory  radio telescopes: the  
R.~C.~Byrd Green Bank Telescope and several Very Long Baseline Array
antennas (supported by the RadioAstron mission''s Green
Bank tracking station, West Virginia, US). The other 3 experiments were performed with the telescopes of
the European VLBI
network (EVN): Effelsberg, Onsala, Wettzell, and the Russian QUASAR
network \cite{quasar-2012-al} stations Svetloe and Zelenchukskaya (supported by the Pushchino tracking station, Moscow region, Russia). Each experiment was made up of a pair of $\sim$~1~hr long sessions separated by $\sim 20$--$30$~hr, which provided a gravitational redshift
modulation between the two sessions of ${\sim (0.5-0.8)\times10^{-10}}$. The observations went
fairly smoothly; the frequency stability of the recorded signal meets our expectations (Fig.~2).

\setlength{\parfillskip}{0pt plus 1fil}

\def\figsubcap#1{\par\noindent\centering\footnotesize(#1)}
\begin{figure}[h]%
\begin{center}
  \parbox{2.1in}{\includegraphics[width=2.15in]{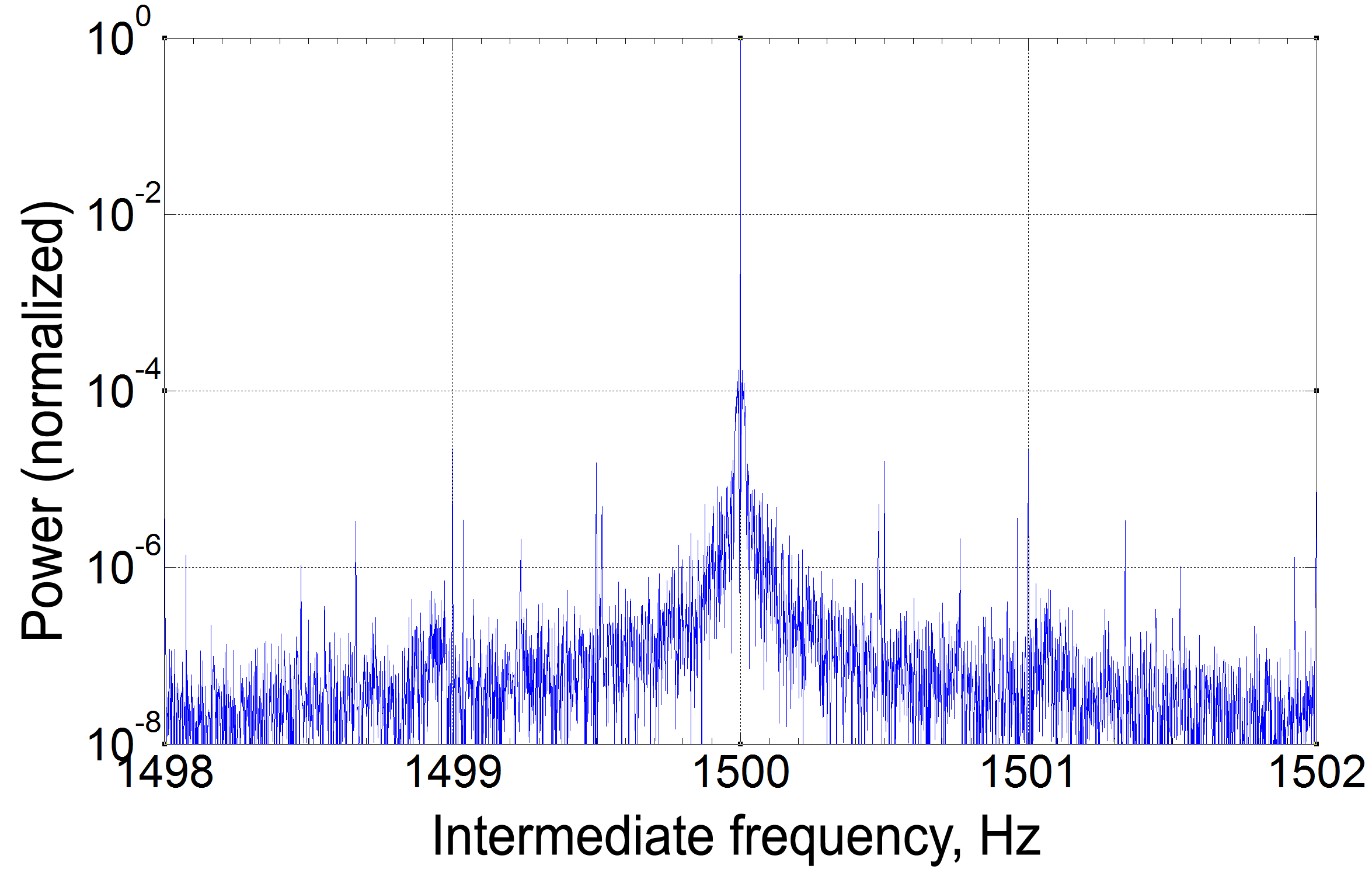}\figsubcap{a}}
  \hspace*{10pt}
  \parbox{2.1in}{\includegraphics[width=2.0in]{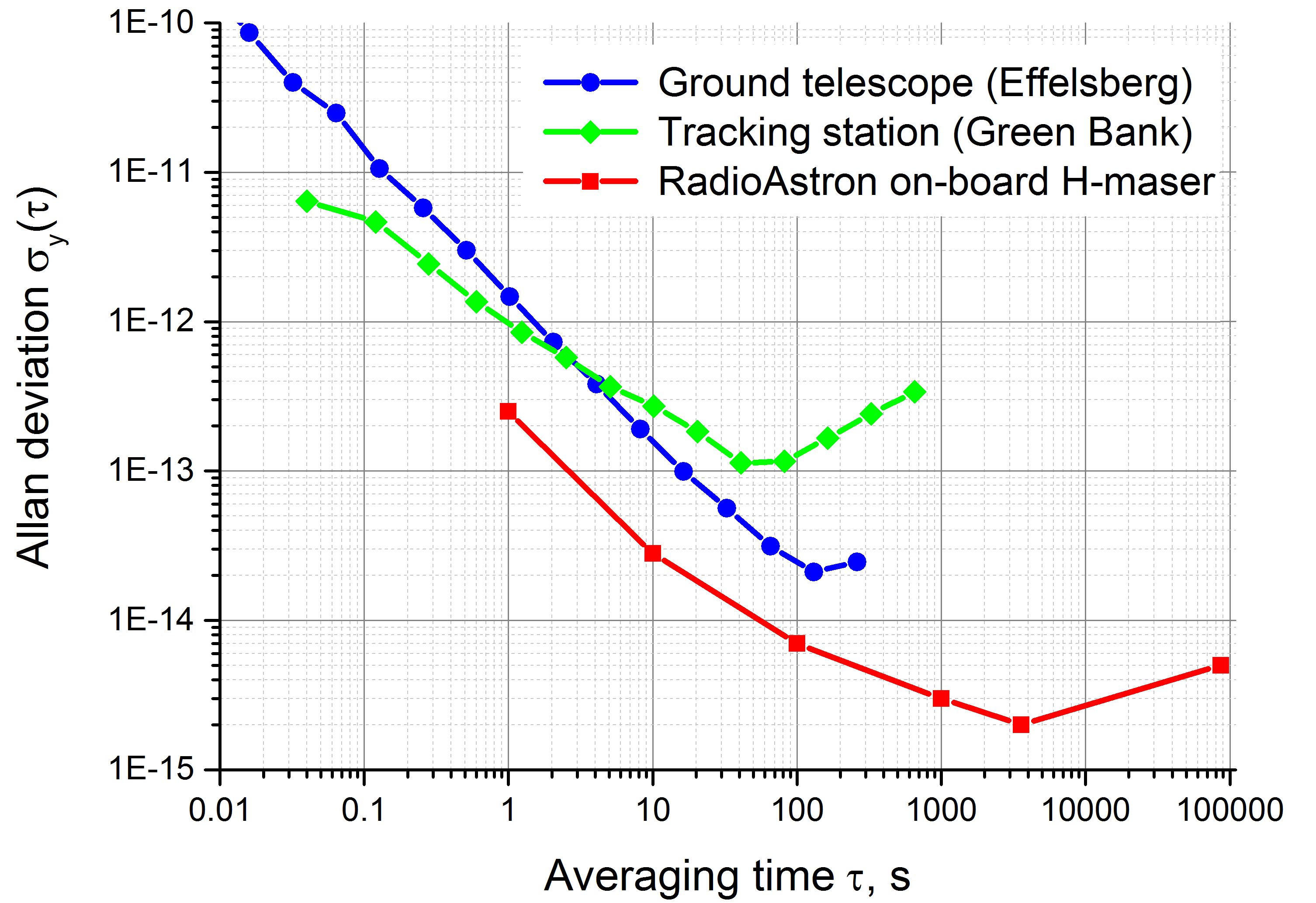}\figsubcap{b}}
  \caption{Frequency stability of the 8.4 GHz RadioAstron downlink signal. (a) Phase-stopped signal spectrum (data
recorded by the Effelsberg telescope, 2015/11/19). (b) Comparison of the
frequency stability
in terms of the Allan deviation of the signal recorded by
a  radio telescope (blue, circles) and a tracking station (green, diamonds).
Allan deviation of the on-board H-maser obtained during laboratory tests (red, squares).}%
  \label{fig1.2}
\end{center}
\end{figure}

The data from
the RadioAstron mission's tracking stations (Pushchino and Green Bank) have been processed and yielded
an accuracy of $\delta\varepsilon\sim4\cdot10^{-4}$.
The data from the radio telescopes, of higher quality 
than tracking
station data (Fig.~2b),\ are currently being processed. We expect to
reach an accuracy of $\sim 8\cdot10^{-5}$ after  processing these data
in full. The most sensitive experiments, i.e.\ with the gravitational redshift effect modulation as large as $3\times10^{-10}$, are planned for the summer of 2016.
The quality of the data
already collected gives a reason to believe that the  accuracy of the gravitational redshift test of $\sim2.5\times10^{-5}$ can be achieved.

\vspace{.3cm}

{\scriptsize The "RadioAstron" project is led by the Astro
Space Center of the Lebedev Physical Institute of
the Russian Academy of Sciences and the Lavochkin
Scientific and Production Association under a contract with the Russian Federal Space Agency, in collaboration with partner organizations in Russia and
other countries.
The European VLBI Network is a joint facility of independent European, African, Asian, and North American radio astronomy institutes. Scientific results from data presented in this publication are derived from the following EVN project code: EL053. The
National Radio Astronomy Observatory is a facility of the National Science
Foundation operated under cooperative agreement by Associated Universities,
Inc.

}

\end{document}